\documentclass
[preprint,pre,nofootinbib,showpacs,titlepage,twocolumn,tightenlines,10pt]{revtex4}%
\usepackage{amsfonts}
\usepackage{amsmath}
\usepackage{amssymb}
\usepackage{graphicx}
\usepackage{acronym}%
\newtheorem{theorem}{Theorem}
\newtheorem{acknowledgement}[theorem]{Acknowledgement}

\begin{document}
\title{Analysis of Nonlinear Synchronization Dynamics of Oscillator Networks by
Laplacian Spectral Methods.}
\author{Patrick N. McGraw and Michael Menzinger}
\affiliation{Department of Chemistry, University of Toronto, Toronto, Ontario, Canada M5S 3H6}

\begin{abstract}
\ We analyze the synchronization dynamics of phase oscillators far from the
synchronization manifold, including the onset of synchronization on scale-free
networks with low and high clustering coefficients. \ We use normal
coordinates and corresponding time-averaged velocities derived from the
Laplacian matrix, which reflects the network's topology. \ \ In terms of these
coordinates, synchronization manifests itself as a contraction of the dynamics
onto progressively lower-dimensional submanifolds of phase space spanned by
Laplacian eigenvectors with lower eigenvalues. \ Differences between high and
low clustering networks can be correlated with features of the Laplacian
spectrum.\ For example, \ the inhibition of full synchoronization at high
clustering is associated with a group of low-lying modes that fail to lock
even at strong coupling, while the advanced partial synchronization at low
coupling noted elsewhere is associated with high-eigenvalue modes. \ 

\end{abstract}
\pacs{89.75.Hc, 05.45.Xt}
\pacs{89.75.Hc, 05.45.Xt}
\pacs{89.75.Hc, 05.45.Xt}
\pacs{89.75.Hc, 05.45.Xt}
\pacs{89.75.Hc, 05.45.Xt}
\pacs{89.75.Hc, 05.45.Xt}
\pacs{89.75.Hc, 05.45.Xt}
\pacs{89.75.Hc, 05.45.Xt}
\pacs{89.75.Hc, 05.45.Xt}
\pacs{89.75.Hc, 05.45.Xt}
\pacs{89.75.Hc, 05.45.Xt}
\pacs{89.75.Hc, 05.45.Xt}
\pacs{89.75.Hc, 05.45.Xt}
\pacs{89.75.Hc, 05.45.Xt}
\pacs{89.75.Hc, 05.45.Xt}
\pacs{89.75.Hc, 05.45.Xt}
\pacs{89.75.Hc, 05.45.Xt}
\pacs{89.75.Hc, 05.45.Xt}
\pacs{89.75.Hc, 05.45.Xt}
\pacs{89.75.Hc, 05.45.Xt}
\pacs{89.75.Hc, 05.45.Xt}
\pacs{89.75.Hc, 05.45.Xt}
\pacs{89.75.Hc, 05.45.Xt}
\pacs{89.75.Hc, 05.45.Xt}
\pacs{89.75.Hc, 05.45.Xt}
\pacs{89.75.Hc, 05.45.Xt}
\pacs{89.75.Hc, 05.45.Xt}
\pacs{89.75.Hc, 05.45.Xt}
\pacs{89.75.Hc, 05.45.Xt}
\pacs{89.75.Hc, 05.45.Xt}
\pacs{89.75.Hc, 05.45.Xt}
\pacs{89.75.Hc, 05.45.Xt}
\pacs{89.75.Hc, 05.45.Xt}
\pacs{89.75.Hc, 05.45.Xt}
\pacs{89.75.Hc, 05.45.Xt}
\pacs{89.75.Hc, 05.45.Xt}
\pacs{89.75.Hc, 05.45.Xt}
\pacs{89.75.Hc, 05.45.Xt}
\pacs{89.75.Hc, 05.45.Xt}
\pacs{89.75.Hc, 05.45.Xt}
\pacs{89.75.Hc, 05.45.Xt}
\pacs{89.75.Hc, 05.45.Xt}
\pacs{89.75.Hc, 05.45.Xt}
\pacs{89.75.Hc, 05.45.Xt}
\pacs{89.75.Hc, 05.45.Xt}
\pacs{89.75.Hc, 05.45.Xt}
\pacs{89.75.Hc, 05.45.Xt}
\pacs{89.75.Hc, 05.45.Xt}
\pacs{89.75.Hc, 05.45.Xt}
\pacs{89.75.Hc, 05.45.Xt}
\pacs{89.75.Hc, 05.45.Xt}
\pacs{89.75.Hc, 05.45.Xt}
\pacs{89.75.Hc, 05.45.Xt}
\pacs{89.75.Hc, 05.45.Xt}
\pacs{89.75.Hc, 05.45.Xt}
\pacs{89.75.Hc, 05.45.Xt}
\pacs{89.75.Hc, 05.45.Xt}
\pacs{89.75.Hc, 05.45.Xt}
\pacs{89.75.Hc, 05.45.Xt}
\pacs{89.75.Hc, 05.45.Xt}
\pacs{89.75.Hc, 05.45.Xt}
\pacs{89.75.Hc, 05.45.Xt}
\pacs{89.75.Hc, 05.45.Xt}
\pacs{89.75.Hc, 05.45.Xt}
\pacs{89.75.Hc, 05.45.Xt}
\pacs{89.75.Hc, 05.45.Xt}
\pacs{89.75.Hc, 05.45.Xt}
\pacs{89.75.Hc, 05.45.Xt}
\pacs{89.75.Hc, 05.45.Xt}
\pacs{89.75.Hc, 05.45.Xt}
\pacs{89.75.Hc, 05.45.Xt}
\pacs{89.75.Hc, 05.45.Xt}
\pacs{89.75.Hc, 05.45.Xt}
\pacs{89.75.Hc, 05.45.Xt}
\pacs{89.75.Hc, 05.45.Xt}
\pacs{89.75.Hc, 05.45.Xt}
\pacs{89.75.Hc, 05.45.Xt}
\pacs{89.75.Hc, 05.45.Xt}
\pacs{89.75.Hc, 05.45.Xt}
\pacs{89.75.Hc, 05.45.Xt}
\pacs{89.75.Hc, 05.45.Xt}
\pacs{89.75.Hc, 05.45.Xt}
\pacs{89.75.Hc, 05.45.Xt}
\pacs{89.75.Hc, 05.45.Xt}
\pacs{89.75.Hc, 05.45.Xt}
\pacs{89.75.Hc, 05.45.Xt}
\pacs{89.75.Hc, 05.45.Xt}
\pacs{89.75.Hc, 05.45.Xt}
\pacs{89.75.Hc, 05.45.Xt}
\pacs{89.75.Hc, 05.45.Xt}
\pacs{89.75.Hc, 05.45.Xt}
\pacs{89.75.Hc, 05.45.Xt}
\pacs{89.75.Hc, 05.45.Xt}
\pacs{89.75.Hc, 05.45.Xt}
\pacs{89.75.Hc, 05.45.Xt}
\pacs{89.75.Hc, 05.45.Xt}
\pacs{89.75.Hc, 05.45.Xt}
\pacs{89.75.Hc, 05.45.Xt}
\pacs{89.75.Hc, 05.45.Xt}
\pacs{89.75.Hc, 05.45.Xt}
\maketitle

The relation between structure and function is a key area in the study of
complex networks \cite{Strogatz}\cite{Netreviews}\cite{Watts}\cite{StructFunc}%
. Synchronization of coupled oscillators\cite{Syncreviews} has applications to
numerous areas of biology including neuroscience, as well as systems such as
coupled lasers and Josephson junctions, and accordingly its dependence on
coupling topology has begun to receive attention. \ \ Among methods of
studying synchronization, the Master Stability Function (MSF)\cite{PecoraMSF}
formalism is appealing because it expresses the dynamical synchronizability in
terms of purely structural features, independent of details of node dynamics.
\ The so-called propensity for synchronization (an indication of the size of
the parameter range giving a stable synchronized state)\cite{Chavez} depends
only on the extremal eigenvalues of the Laplacian matrix. \ Within this
formalism, the effects of small-world properties, heterogeneity and certain
types of weighted coupling have been examined.\cite{MSFcomparisons}%
\cite{Chavez} \ The MSF, however, is restricted to the linear domain, close to
exact amplitude and phase synchronization of chaotic oscillators. \ Others
\cite{Moreno}\cite{Hong}\cite{Ichinomia}\cite{McGraw}\cite{Zaragoza} have
examined numerically and analytically the onset of synchronization for phase
oscillators coupled on networks, a problem for which the MSF is unsuited. \ 

In this report we demonstrate an application of the Laplacian spectrum to a
sparsely connected network Kuramoto \cite{Kuramoto} model both close to and
far from full synchronization. \ As a case study, we examine scale-free
networks with low and high clustering coefficients, examined elsewhere by
different methods\cite{McGraw}. \ \ Parametrizing the phase space with normal
coordinates based on Laplacian eigenvectors, \ we show in these two sample
cases that with increasing coupling strength, \ the dynamics contracts onto
progressively lower-dimensional subspaces spanned by lower-lying (less stable)
eigenvectors. \ \ Dynamical properties of the networks can be correlated with
specific features of their spectra. \ By focusing on appropriately chosen
collective degrees of freedom (the normal coordinates), our approach
complements methods of analysis that focus on the locking and unlocking of
individual oscillators\cite{Kuramoto}\cite{McGraw}. \ In the spirit of the
MSF, our analysis highlights the effects of network topology via the spectrum,
but in contrast it applies to a range of desynchronized and partly
synchronized states, not only to incipient deviations from full
synchronization. \ We consider the the spectrum in its entirety, not only the
extremal eigenvalues. \ The coordinates derived from the Laplacian spectrum
provide a helpful empirical tool for the analysis of simulation results. \ We
use them here to gain new insight into the different behaviors of networks
with high and low clustering coefficients. \ Our emphasis is on the process of
synchronization, rather than on rigorous bounds for the threshold of
desynchronization. \ 

We first define the model and show how the Laplacian and its spectrum appear
naturally in a linearized description of the frequency-synchronized state.
\ Then we use the Laplacian eigenvectors to parametrize the partially
desynchronized states and show that this coordinate system remains useful well
beyond the range of validity of the linearization. \ 

Our model\cite{Moreno}\cite{Hong} is defined by the coupled equations%
\begin{equation}
\frac{d\phi_{i}}{dt}=\omega_{i}+\frac{\beta}{\left\langle k\right\rangle }%
{\displaystyle\sum\limits_{j}}
a_{ij}\sin(\phi_{i}-\phi_{j}),\label{Kuramotomodel}%
\end{equation}
where $\phi_{i}$ are $N$ phase variables (one associated with each node of a
network), \ $-1\leq\omega_{i}\leq1$ are the randomly and uniformly distributed
intrinsic frequencies\footnote{The average of $\omega_{i}$ can be taken to be
$\overline{\omega}=0$ without loss of generality (if it is not zero it can be
made so by changing variables into a rotating frame of reference.)\ }, $\beta$
is the overall coupling strength, and $a_{ij}$ is the weighting matrix of the
individual couplings. \ \ In our examples, all links are weighted equally, and
$a_{ij}$ is simply the adjacency matrix ($a_{ij}=1$ if $i$ and $j$ are
connected, 0 otherwise). As in \cite{Hong} and \cite{McGraw} the coupling
strength is normalized by the average degree $\left\langle k\right\rangle $ of
all nodes.\footnote{By normalizing the average total input to a unit this
convention facilitates comparisons among networks and corresponds to the
normalization by $N$ in the original fully connected Kuramoto model. \ }
\ \ \ At low coupling strength, each oscillator moves independently at its
intrinsic frequency, \ but as the coupling increases some become mutually
entrained. \ At sufficiently strong coupling, \ all oscillators rotate at the
same frequency, $\frac{d\phi_{i}}{dt}=\overline{\omega}=0$. \ For the
original, fully connected model the steady-state phases depend only on the
intrinsic frequencies $\omega_{i}$: \ those with higher frequencies lead the
ensemble while those with lower frequencies lag. \ In the present sparsely
connected version, on the other hand, \ each oscillator is influenced
differently by its local neighborhood. \ \ If the phase differences are small
then the sine coupling function can be approximated linearly and\ the
equations of motion become%
\begin{equation}
\frac{d\phi_{i}}{dt}=\omega_{i}+\frac{\beta}{\left\langle k\right\rangle }%
{\displaystyle\sum\limits_{j}}
a_{ij}(\phi_{i}-\phi_{j})=\omega_{i}-\frac{\beta}{\left\langle k\right\rangle
}%
{\displaystyle\sum\limits_{j}}
\pounds _{ij}\phi_{j}\label{linearEOM}%
\end{equation}
where
\begin{equation}
\pounds _{ij}=a_{ij}-\delta_{ij}%
{\displaystyle\sum\limits_{k}}
a_{ik}=a_{ij}-\delta_{ij}k_{i}\label{LaplacianDef}%
\end{equation}
\ is the Laplacian matrix. \ The degree $k_{i}$ of the $i$th node is defined
as the number of nodes to which it is connected, and the second equation in
(\ref{LaplacianDef}) thus holds if the couplings are equally weighted. \ \ 

The steady-state (frequency locked) phases can be found by diagonalizing the
Laplacian. \ \ Let its normalized eigenvectors and corresponding eigenvalues
be $\mathbf{v}^{\alpha}$ and $\lambda^{\alpha}$, where $1\leq\alpha\leq N$.
\ Enumerating lattice sites by Latin indices and Laplacian eigenvectors by
Greek ones, we define projections of the phase and frequency vectors onto
these eigenvectors by
\begin{equation}
\phi^{\alpha}\equiv\sum_{i}\phi_{i}v_{i}^{\alpha},\;\omega^{\alpha}\equiv
\sum_{i}\omega_{i}v_{i}^{\alpha}, \label{NormalCoord}%
\end{equation}
which allows the equations of motion (\ref{linearEOM}) to be rewritten as%
\begin{equation}
\frac{d\phi^{\alpha}}{dt}=\omega^{\alpha}\ -\frac{\beta}{\left\langle
k\right\rangle }\lambda^{\alpha}\phi^{\alpha}. \label{DiagonalEOM}%
\end{equation}
The steady state values $\overline{\phi^{\alpha}}$ of the normal coordinates
are given by \
\begin{equation}
\overline{\phi^{\alpha}}=\frac{\langle k\rangle}{\beta\lambda^{\alpha}}%
\omega^{\alpha}. \label{steadyphase}%
\end{equation}
\ Relaxation to this equilibrium obeys
\begin{equation}
\frac{dx^{\alpha}}{dt}=-\frac{\beta\lambda^{\alpha}}{\langle k\rangle
}x^{\alpha} \label{relaxation}%
\end{equation}
where $x^{\alpha}=\phi^{\alpha}-\overline{\phi^{\alpha}}$ is the displacement
from equilibrium along the $\alpha$-th normal coordinate. \ \ The equilibrium
is stable provided all $\lambda^{\alpha}\geq0$ and the phase displacements are
small enough for the linear approximation to hold. \ By the definition
(\ref{LaplacianDef}), the row sum $\sum_{j}\pounds _{ij}$ of the Laplacian is
zero for all rows and therefore $(1,1,...1)$ is always an eigenvector with
eigenvalue $0$, \ but for a connected network, \ all other eigenvalues are
positive\cite{Mohar}. \ Therefore, the frequency synchronized state is
neutrally stable against a uniform shift of all phases, but stable against all
other perturbations. \ The stability breaks down only due to nonlinear
effects: \ \ the slope of the sinusoidal coupling function decreases with
increasing phase differences and eventually ceases to provide sufficient
restoring force. \ \ \ Since the phase displacements are largest along the
eigenvectors with lowest $\lambda^{\alpha}$, \ \ these eigenvectors represent
modes along which frequency synchronization first fails as the coupling
decreases. \
\begin{figure}
[ptb]
\begin{center}
\includegraphics[
height=1.9726in,
width=2.4474in
]%
{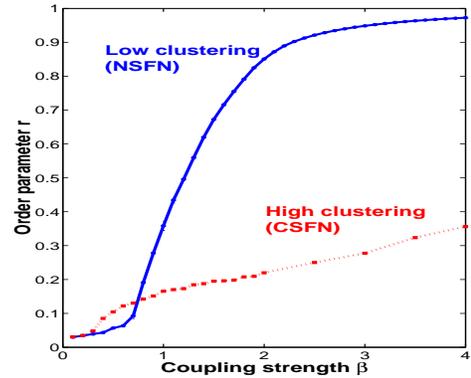}%
\caption{Synchronization order parameter as a function of coupling strength
for the two scale-free networks. \ The strongly clustered network (CSFN)
undergoes a partial synchronization at lower coupling strength, but at higher
couplings it is significantly less synchronized than the normal (NSFN)
network. \ }%
\label{orderparamfig}%
\end{center}
\end{figure}
\ \ 

Although they arise most naturally from the linear analysis, the Laplacian
eigenvectors retain their usefulness beyond that approximation. \ To
demonstrate this, \ we consider two networks as examples. \ Our two networks
have identical, scale-free, degree distributions but differ in their
clustering coefficient \cite{Watts}\cite{WS} --- a measure of the likelihood
that two neighbors of a given node are also directly connected to each other,
or a measure of the prevalence of triangles in the network topology. \ The
first is a Barabasi-Albert \cite{BA} scale-free network of $N=1000$ nodes with
average degree $\left\langle k\right\rangle =20$, \ grown by means of
preferential attachment beginning with a fully connected core of $m=10$ nodes.
\ The Barabasi-Albert network has a low clustering coefficient,
\ approximately 0.02. \ \ The other network is derived from the first by
applying Kim's \cite{Kim} stochastic rewiring method to increase the
clustering coefficient to 0.62, \ without changing the degree distribution
(although, as mentioned below, some other properties vary in tandem with the
clustering). \ \ We will refer to these networks as the normal scale-free
network (NSFN) and the clustered scale-free network (CSFN) respectively.
\ These were among the networks studied previously in \cite{McGraw}, \ where
it was found that increased clustering inhibited full synchronization at high
$\beta$ but surprisingly promoted the onset of partial synchronization at low
$\beta$. \ This behavior is shown in a plot (Fig. 1) of the standard
synchronization order parameter
\begin{equation}
r=\left\langle \left\vert \sum_{j}e^{i\phi_{j}}\right\vert \right\rangle _{T}
\label{orderparam}%
\end{equation}
(where $\left\langle ..\right\rangle _{T}$ stands for a time average) as a
function of the coupling strength. \ \ In the unsynchronized state at low
coupling, \ $r=O(1/\sqrt{N})$ from both networks. \ The onset of
synchronization is shown by an upward turn in the plot of $r$ vs. $\beta$.
\ This transition occurs at a lower $\beta$ for the CSFN than for the NSFN, so
that for $0.25\lesssim\beta\lesssim0.75$, \ $r$ is larger for the CSFN. \ At
higher couplings, however, \ the CSFN strongly resists full synchronization
and remains in a partly synchronized state with a much smaller value of $r$
than for the NSFN. \ \ 

The Laplacian eigenvalue spectra of the NSFN and CSFN are shown in figure 2.
\ Like the degree distribution, the distribution of eigenvalues has a
power-law tail in both cases. \ An important difference appears at the lower
end of the spectrum. \ In the NSFN, there is a gap between the lowest nonzero
eigenvalue and zero, \ and the single peak of the distribution is near this
lower cut-off. \ \ The CSFN spectrum, on the other hand, has a second peak
close to zero, indicating a number of nearly degenerate quasi-zero modes.
\ \ The presence of eigenvalues close to zero indicates that the network has a
strong community structure,\cite{Capocci} i.e., \ it consists of components
(communities) that have fewer connections between different components than
within each component. \ \ In fact, the low eigenvectors of the Laplacian form
the basis of some algorithms for detecting communities \cite{Donetti}%
\cite{Capocci}\ \ From the spectrum, then, we learn that the rewiring
algorithm has not only created clustering (a local property measuring the
number of triangles) but as a byproduct has also created global communities.
\ Since higher clustering means more "local" connections at the expense of
\ long-range ones, this is not surprising, but neither is it inevitable--- for
example, a regular ring or a "small-world" network of the type considered in
\cite{WS} \ has high clustering but no communities.%
\begin{figure}
[ptb]
\begin{center}
\includegraphics[
height=2.5953in,
width=2.2969in
]%
{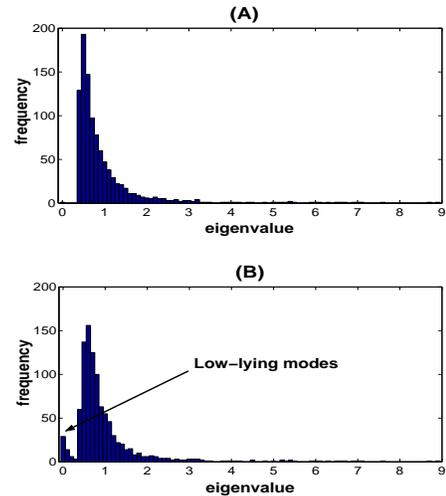}%
\caption{Histograms of the scaled Laplacian eigenvalues $\lambda^{\alpha
}/\left\langle k\right\rangle $ for the NSFN (A) and CSFN (B). \ Both
histograms have power-law tails at large eigenvalues. \ A key difference is
the group of low-lying modes, separate from the main spectrum, in the highly
clustered network. }%
\label{spec}%
\end{center}
\end{figure}
\ 

\ To further aid in analyzing the dynamics we define the observed frequencies
(rotation numbers) of the oscillators as the time averages%
\begin{equation}
\Omega_{j}=\left\langle \frac{d\varphi_{j}}{dt}\right\rangle _{T}.
\label{obsfreqdef}%
\end{equation}
Projecting the vector of observed frequencies onto the Laplacian eigenbasis
gives a time-averaged velocity along the direction defined by each
eigenvector:%
\begin{equation}
\Omega^{\alpha}=\sum_{j}\Omega_{j}v_{j}^{\alpha}. \label{velocitydef}%
\end{equation}
\ \ \ In a fully frequency-synchronized state, \ $\Omega^{\alpha}=0$ for all
$\alpha$.\ \ \ In figure 3 ensemble averages of the squares of the velocities
\ $\left[  (\Omega^{\alpha})^{2}\right]  _{\omega}$ \ are plotted against the
eigenvalues $\lambda^{\alpha}$ for both networks at three values of the
coupling strength. \ \ The average $\left[  ..\right]  _{\omega}$ is over 25
different realizations of the random frequency distribution. \ \ \ In a case
where the network is almost completely incoherent (for example, the NSFN at
$\beta=0.5$), \ all velocities $\Omega^{\alpha}$ are random and of
approximately equal magnitude. \ \ In the case of full synchronization (NSFN
at $\beta=2.5$), \ all velocities are \textquotedblleft
locked\textquotedblright\ at zero. \ \ \ At intermediate values af $\beta$,
however, some modes are locked while others are "drifting" at nonzero
velocities. \ \ It is clear from the plots that as the coupling increases,
modes with higher eigenvalues lock sooner than those with lower ones ---
synchronization proceeds from the top of the spectrum downward.
\ Synchronization manifests itself as a progressive contraction of the
dynamics onto lower-dimensional submanifolds of the phase space.%
\begin{figure}
[ptb]
\begin{center}
\includegraphics[
height=3.9565in,
width=2.994in
]%
{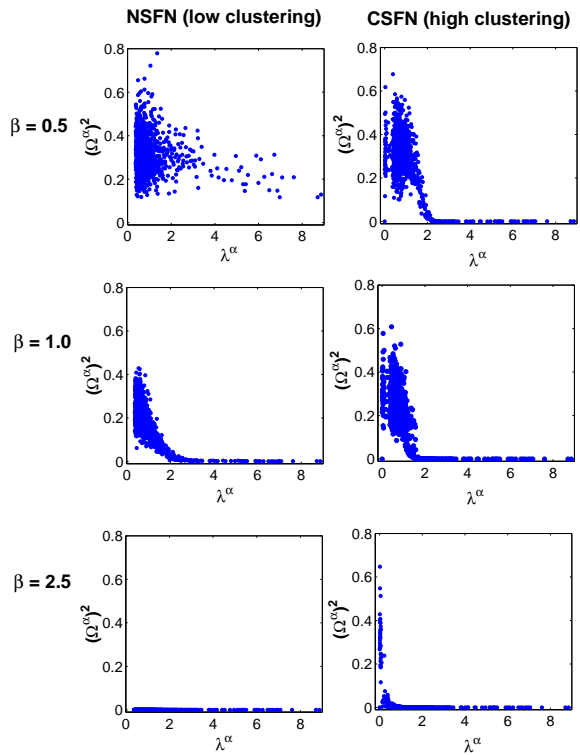}%
\caption{Mean square projections of observed frequency differences onto
Laplacian eigenvectors (normal velocities) at several values of coupling
strength $\beta$. \ Each point represents an average over 25 realizations of
the random intrinsic frequencies. \ In both types of networks, \ velocities
along the eigenvectors with higher eigenvalues vanish at lower couplings than
those with lower eigenvalues. \ In the highly clustered network (right
column), two features are notable: \ The higher mode velocities vanish more
readily than for the low-clustering network, and the lowest lying modes
maintain nonzero velocities even at $\beta=2.5$ where all others vanish. \ }%
\label{velocities}%
\end{center}
\end{figure}
\ 

In the case of the CSFN, \ higher modes begin to lock more readily than in the
NSFN, indicating that these high modes in the spectrum are implicated in the
advanced partial synchronization of the clustered network (fig.
\ref{orderparamfig}). \ \ At stronger coupling, on the other hand, \ the most
notable difference of the clustered from the normal network is that the
low-lying modes associated with community divisions (fig. \ref{spec}) continue
to drift while all others are locked. \ The lack of synchronization \ is
associated with these low-lying modes, and the frequency clusters noted in
this case \cite{McGraw}\ coincide with topological communities. \ \ The
observation that these low-lying modes fail to lock is consistent with our
intuition based on the linear approximation, according to which these modes
represent the strongest potential instabilities of a synchronized state.
\ Their presence in the spectrum accounts for the inhibition of full
synchronization in the CSFN. \ \ The finding that different sets of
eigenvectors are involved in the two regimes supports the claim
\cite{Zaragoza} that two separate effects are at work, with the advanced onset
being an effect of the clustering per se, which is a local property, while the
delay of full synchronization results from global topological properties that
are correlated with clustering. \ \ In particular, the delay was ascribed to
effects of increasing average path length\cite{Zaragoza}. \ \ However, the
involvement of the low eigenvectors associated with communities and the
dymanical fragmentation of the network into synchronized subgroups suggest
that it is more specifically a function of the community structure (although
the latter certainly is correlated with a long average path length).
\ \ Ongoing studies aim to further disentangle the various correlated
topological features and their effects.

Examining the dynamics in terms of normal coordinates defined by the Laplacian
eigenvectors provides a geometric basis for viewing the flow of the ensemble
of oscillators that complements other tools of analysis such as global order
parameters\cite{Kuramoto} or scatter plots of observed vs. intrinsic
frequencies of individual oscillators\cite{McGraw}. \ Like the MSF formalism,
it gives a\ partial picture of how purely structural features influence the
synchronization dynamics, since the Laplacian reflects only the network
topology. \ It is not obvious \emph{a priori} that Laplacian eigenvectors
should be relevant beyond the range of validity of the linear approximation
near a fully phase-synchronized state, \ yet the normal coordinate velocities,
in particular, contain nontrivial dynamical information well away from this
limit, and they split into subsets associated with different dynamical effect.
\ They allow one to indentify collective degrees of freedom responsible for on
one hand the advanced partial synchronization and on the other the inhibition
of complete synchronization in a highly clustered scale-free network.
\ Connections among topology, spectrum and dynamics will be explored more
fully in a future publication, which will apply the formalism to other types
of networks including ones with unequal and asymmetric couplings $a_{ij}$,
\ as well as considering other spectral properties such as the localization
and delocalization of modes. \ 

\begin{acknowledgement}
This work was supported by the NSERC of Canada.
\end{acknowledgement}


\bigskip

\begin{thebibliography}{99}                                                                                               %


\bibitem {Strogatz}\ S.H. Strogatz, Nature \textbf{410}, 268 (2001).

\bibitem {Netreviews}R. Albert and A.L. Barabasi, Rev. Mod. Phys. \textbf{74},
47 (2002); \ M.E.J. Newman, preprint cond-mat/0202208 (2002); \ A.L. Barabasi,
\textit{Linked: The New Science of Networks} (Perseus Publishing, Cambridge,
MA, 2002); \ S.N. Dorogovtsev and J.F.F. Mendes, \textit{Evolution of
Networks: From Biological Nets to the Internet and WWW} (Oxford University
Press, Oxford, 2003). \ 

\bibitem {Watts}D.J. Watts, \textit{Small Worlds} (Princeton University Press,
Princeton, NJ, 1999). \textit{Six Degrees }(Norton, New York, 2003\textit{).}

\bibitem {StructFunc}M.E.J. Newman, SIAM Review \textbf{45}, 167 (2003); \ S.
Boccaletti, V. Latora, Y. Moreno, M. Chavez and D.-U. Hwang, \ Phys. Rep.
\textbf{424}, 175 (2006)

\bibitem {Syncreviews}S.H. Strogatz, \textit{Sync: The Emerging Science of
Spontanteous Order} (Hyperion, New York, 2003). \ 

\bibitem {PecoraMSF}L. M. Pecora and T.L. Carroll, Phys. Rev. Lett.
\textbf{80}, 2109 (1998).

\bibitem {Chavez}M. Chavez, D.U. Hwang, A. Amann, \ H.G.E. Hentcshel and S.
Bocaletti, Phys. Rev. Lett. \textbf{94}, 218701 (2005). \ 

\bibitem {MSFcomparisons}M. Barahona and L.M. Pecora, Phys. Rev. Lett.
\textbf{89}, 054101 \ (2002); \ T. Nishikawa, A.E. Motter, Y.-C. Lai and F.C.
Hoppensteadt, Phys. Rev. Lett.\textbf{ 91}, 014101 \ (2003).

\bibitem {Moreno}Y. Moreno and A.F. Pacheco, Europhys. Lett. \textbf{68}, 603 (2004).

\bibitem {Hong}H. Hong, M.Y. Choi and B.J. Kim, Phys. Rev E \textbf{65},
026139 (2002).

\bibitem {Ichinomia}T. Ichinomiya, Phys. Rev. E \textbf{70}, 026116 (2004). \ 

\bibitem {McGraw}P. N. McGraw and M. Menzinger, Phys. Rev. E \textbf{72},
026210 (2005).

\bibitem {Zaragoza}J. Gomez-Gardenes and Y. Moreno, \ preprint
cond-mat/0608309 \ (2006). \ 

\bibitem {Kuramoto}Y. Kuramoto, \textit{Chemical oscillations, waves and
turbulence }(Springer Verlag, \ Berlin; 1984) ; A. Pikovsky,
\textit{Synchronization, }(Cambridge University Press, Cambridge; 2001).

\bibitem {Mohar}B. Mohar, "The Laplacian Spectrum of Graphs," in Y.Alavi, G.
Chartrand, O.R. Ollermann, A.J. Schwenk, ed., \textit{Graph Theory,
Combinatorics and Applications} (Wiley, New York), 871-98 (1991). \ 

\bibitem {WS}D.J. Watts and S.H. Strogatz, Nature \textbf{393}, 440 (1998).

\bibitem {BA}A.L. Barabasi and R. Albert, Science \textbf{286}, 509 (1999).

\bibitem {Kim}B.J. Kim, Phys. Rev. E, \textbf{69}, 045101(R) (2004).

\bibitem {Donetti}L. Donetti and M.A. Mu\~{n}oz, J.Stat. Mech, P10012 (2004). \ 

\bibitem {Capocci}A. Capocci, V.D.P. Servedio, G. Caldarelli and F. Colaiori,
Physica A \textbf{352}, 669 (2005).
\end{thebibliography}
\end{document}